\documentclass{llncs}

\usepackage{makeidx}  
\begin{document}
%
%
\pagestyle{headings}  
%
% ...
\mainmatter              
\title{A brief overview of the Pawns\\
programming language}
\titlerunning{Pawns language}  
%
\author{Lee Naish}
\authorrunning{L. Naish}   
%
\tocauthor{Lee Naish}
\institute{University of Melbourne, Melbourne 3010, Australia\\
\email{dr.lee.naish@gmail.com},\\
\texttt{https://lee-naish.github.io/} }

\maketitle              

\begin{center}
\today
\end{center}
\begin{abstract}
Pawns is a programming language under development which supports pure
functional programming (including algebraic data types, higher order
programming and parametric polymorphism) and imperative programming
(including pointers, destructive update of shared data structures and
global variables), integrated so each can call the other and with purity
checked by the compiler.  For pure functional code the programmer need not
understand the representation of the data structures.  For imperative code
the representation must be understood and all effects and dependencies
must be documented in the code. For example, if a function may update one
of its arguments, this must be declared in the function type signature
and noted where the function is called. A single update operation may
affect several variables due to sharing of representations (pointer
aliasing).  Pawns code requires all affected variables to be annotated
wherever they may be updated and information about sharing to be declared.
Annotations are also required where IO or other global variables are used
and this must be declared in type signatures as well.  Sharing analysis,
performed by the compiler, is the key to many aspects of Pawns. It
enables us to check that all effects are made obvious in the source
code, effects can be encapsulated inside a pure interface and effects
can be used safely in the presence of polymorphism.

\noindent
Keywords: functional programming language, destructive update, mutability,
effects, algebraic data type, sharing analysis
\end{abstract}
\section{Introduction}
\label{sec-intro}


This paper briefly describes the main features Pawns, a programming
language that is currently under development.  The aim is to convey
a feel for the general ideas; \cite{pawns-intro} does the same but
includes significantly more detail, discussion of language design
issues and citation of related work.  We assume the reader is familiar
with Haskell and C.  Pawns supports pure functional programming with
strict evaluation, algebraic data types, parametric polymorphism,
and higher order programming.  It also supports ``impure" code, such
using state (including IO) and destructive update of all compound data
types via pointers (references or ``refs" for short) but all such
code is highlighted by ``!" annotations.  A call to a function that
relies on state must be prefixed by ``!"; the details of the state(s)
are declared in the type signature.  Additionally, variables that are
updated must be prefixed with ``!".  A function call with no ``!" is
guaranteed to behave as a pure function, though Pawns allows impurity
to be encapsulated (and checked by the compiler), so the function
may be implemented using impure features.  The representations of
different variables can be shared, so updating one variable may also
update other variables and the Pawns compiler checks that all relevant
variables are annotated with ``!" at that point in the source code:
Pawns is an acronym for ``Pointer assignment without nasty surprises"
and its most important (and complex) innovation is the way update of
shared data structures is supported and how pure and impure code can
be mixed.  Impure programming in Pawns can be like programming in C,
with destructive update of fields of structs representing ADT values
and performance equal to or better than portable C. However, there no
unsafe operations (such as dereferencing possibly NULL pointers, casts,
acessing fields of unions, etc) and all interractions/dependencies due
to sharing must be documented in annotations/declarations.

The rest of this paper is structured as follows.
Section \ref{sec-pure} gives a simple example of pure functional
programming.
Section \ref{sec-rep-du} describes how destructive update is done in
Pawns and gives some information about data representation.
Section \ref{sec-du-eg} gives a two examples of code using destructive
update in Pawns, mentioning sharing of data structures but deferring the
details of how sharing is handled.
Section \ref{sec-pure-abs} discusses the distinction between data
structures that can be simply viewed as abstract values (typical in pure
code) and those for which sharing must be understood (a necessity when
destructive update is used).
Section \ref{sec-sharing} discusses how sharing and destructive update
information is incorporated into Pawns type signatures and the kind of
sharing analysis done by the compiler.
Section \ref{sec-state} presents how IO and other forms of ``state'' can
be used in Pawns.
Section \ref{sec-rename} discusses a Pawns feature that allows renaming of
functions so different type signatures can be given, overcoming some of
limitations of polymorphism, particularly for impure Pawns code.
Section \ref{sec-comp} briefly discusses some of the additional complications
surrounding safety in Pawns.
Section \ref{sec-conc} concludes.



\section{Pure functional programming example - BST creation}
\label{sec-pure}

Consider the task of converting a list of integers into a binary search
tree.  Pawns supports typical pure functional programming solutions
such as Figure \ref{fig_list_bst_pure}, presented using Haskell-like
syntax\footnote{Pawns currently only supports a temporary syntax, to
avoid decisions on syntax and the need to write a parser}. Note the use
of polymorphic algebraic data types and the polymorphic higher order
function \texttt{foldl}; Pawns does not currently support type classes
or existential types.

\begin{figure}
\begin{verbatim}
-- polymorhic List type (actually built in)
data List t = Nil | Cons t (List t)
type Ints = List Int
data BST = Empty | Node BST Int BST

-- convert list of integers to BST (pure code)
list_bst_pure:: Ints -> BST
list_bst_pure xs =
    foldl bst_insert_pure Empty xs

-- insert integer into a BST to give new BST
-- (pure; re-builds a path from root to a leaf)
bst_insert_pure:: BST -> Int -> BST
bst_insert_pure t0 x =
    case t0 of
    Empty ->
        Node Empty x Empty
    (Node l n r) ->
        if x <= n then
            Node (bst_insert_pure l x) n r
        else
            Node l n (bst_insert_pure r x)

-- standard library foldl for lists
foldl:: (b -> a -> b) -> b -> List a -> b
foldl f y xs =
    case xs of
    Nil ->
        y
    (Cons x xs1) ->
        foldl f (f y x) xs1
\end{verbatim}
\caption{BST creation using pure code}
\label{fig_list_bst_pure}
\end{figure}

An advantage of this style of programming that it is not necessary to
understand how values are represented in order to write and reason about
the code. However, \verb@bst_insert_pure@ builds a new node at
each level of the tree visited so it is much less efficient (a factor of
around twenty in our experiments) than just using destructive update
when a leaf is reached. BST creation is unlikely to be a major time
component of any application but we will use this as a simple example
of how destructive update can be used in Pawns.

\section{Representation and destructive update of values}
\label{sec-rep-du}

The key thing to note about data representation and update in Pawns is
that \emph{arguments of data constructors are stored in main memory} and
these are the only things that can be updated. The data constructors
themselves are like pointers (they may be a pointer plus a ``tag"
or, for data constructors with no arguments, they may just be a small
integer). The list \texttt{(Cons 42 Nil)} is represented as a pointer
to two memory cells, containing the integer 42 and a small number that
represents \texttt{Nil}, respectively - the same as a linked list in
C. Similarly, a BST is represented essentially using pointers to structs
with three fields. For types that have more than one data constructor
with arguments (such as the cord data structure discussed in Section
\ref{sec-du-eg}) the representation uses tags and is more efficient than
portable C code; see \cite{adtpp} for details. Pawns allows the kind
of programming we can do in C with pointers to structs and assignment
to fields of structs. There is also additional flexibility because an
ADT can have any number of data constructors with arguments (which is
like having a pointer to any number of different struct types) and any
number of data constructors with no arguments (like have any number of
different NULL values) and all operations are safe (no dereferencing of
NULL values, no casts, \emph{et cetera}).

Pawns variables are not names for memory locations that can be updated ---
it is not possible to assign to an existing variable or get a ``pointer
to a variable'' as you can in C. However, the representation of the value
of a variable may have mutable components.  For example, a variable whose
value is \texttt{(Cons 42 Nil)} will always be a \texttt{Cons} pointer to
the same two memory cells but the content of these cells can potentially
be updated, changing the overall value of the variable.  All update
is done via special pointer (ref) types (similar to \texttt{STRef} in
Haskell and \texttt{ref} in ML).  There is a polymorphic \texttt{Ref
t} type that is a pointer to a memory cell containing a value of type
\texttt{t}. You can think of the memory cell as the argument of the
data constructor for the \texttt{Ref} type, thus it can be updated.
However, Pawns code never uses an explicit data constructor for refs
but instead just uses a dereference operator, ``\verb@*@", like C. If
\texttt{x} is a Pawns expression of type \texttt{Ref t} then \verb@*x@
is the value of type \texttt{t} that \texttt{x} points to. There are no
\texttt{NULL} refs.

The simplest way to create a ref is by using a let binding with \texttt{*}
prefixing the let-bound variable. The ``let'' and ``in'' keywords of
Haskell are not required in Pawns and ``;" is used for sequencing, thus
\texttt{x = 42; *xp = 42} creates two variables, the first of which equals
42 and the second points to a newly allocated memory cell containing 42
(similar to the Haskell monadic code \texttt{x <- newSTRef 42}, or an
ML let expression with \texttt{x = ref 42}).  Destructive update is done
by dereferencing a pointer on the left of the ``\verb@:=@" (assignment)
operator. All variables\footnote{More precicely, all live variables;
those which are never used again can generally be ignored.} that are affected
must be prefixed by ``!". Typically there will be a pointer variable on
the left (so \verb@*xp := ...@ is written \verb@*!xp := ...@) but there
may also be other variables that share its representation; these can be
annotated with ``!" at the right of the statement. Figure \ref{fig_du}
has a simple example.

\begin{figure}
\begin{verbatim}
    x = 42;         -- let binding of x to 42
    *xp = x;        -- xp points to a new memory cell containing 42
    yp = xp;        -- yp points to the same memory cell
    y = *yp;        -- y is the contents of the memory cell (42)
    *!xp := 43 !yp; -- update what xp points to (also affects yp!)
    z = *yp         -- z is the contents of the memory cell (43)
\end{verbatim}
\caption{Destructive update via a ref}
\label{fig_du}
\end{figure}

Without the ``\texttt{!y}" annotation, both \texttt{y} and \texttt{z}
would be bound to \verb@*yp@ with no intervening occurrence of \texttt{yp}
in the code, yet they end up with different values. This is typical
of the potentially confusing ``surprises'' encountered in languages
that support code for destructive update with pointer aliasing and
shared data structures, which is needed for many important algorithms.
Pawns supports such code but insist the programmer documents sharing
and effects, in a way that can be checked by the compiler.

Just as prefixing a variable with \verb@*@ in a let binding creates a
pointer variable, the same can be done with pattern bindings. These
``dereference patterns" are an important innovation of Pawns.
For example, the code for \verb@bst_insert_pure@ could be rewritten
as in Figure \ref{fig_bst_insert_pure_p}.  Instead of the pattern
matching with a \texttt{Node} creating variables of type BST and Int,
it creates variables of type \texttt{Ref BST} and \texttt{Ref Int},
which are pointers to the arguments of the \texttt{Node} data constructor.
Refs are created but no extra memory cells are allocated and no monads or
changes to the \texttt{BST} type are required; there is no equivalent in
languages such as Haskell and ML.  The subsequent code simply dereferences
the pointers to obtain the same values as before and the code is pure ---
refs/pointers themselves do not introduce impurity. However, such pointers
could potentially be used to destructively update the \texttt{Node}
arguments (which is impure).

\begin{figure}
\begin{verbatim}
bst_insert_pure_p t0 x =
    case t0 of
    Empty ->
        Node Empty x Empty
    (Node *lp *np *rp) ->    -- creates refs/pointers to Node arguments
        if x <= *np then
            Node (bst_insert_pure_p *lp x) *np *rp
        else
            Node *lp *np (bst_insert_pure_p *rp x)
\end{verbatim}
\caption{BST insertion using pure code with pointers}
\label{fig_bst_insert_pure_p}
\end{figure}

\section{Destructive update examples}
\label{sec-du-eg}

We now give two short examples of using destructive update in Pawns.
The first is an alternative way to construct a BST and the second
is an example where the sharing of data structures is more complex.
Building a BST from a list of integers can be done very efficiently by
first allocating a memory cell containing an empty BST then repeatedly
traversing down the tree and destructively inserting the next integer as
a new leaf  --- see Figure \ref{fig_list_bst_du}.  Both \verb@foldl_du@
and \verb@bst_insert_du@ simply return void because the tree is updated
in situ but because of the destructive update (they are not pure
functions), Pawns insists more information is provided in their type
signatures; we will discuss this in Section \ref{sec-sharing}. However,
\verb@list_bst_du@ behaves as a pure function, indistinguishable from
\verb@list_bst_pure@, even though it is defined in terms of impure
functions (and is far more efficient). To construct the BST it is
necessary to consider low level details such as the representation
of the tree and any sharing present but after it is returned from
\verb@list_bst_du@ it can be treated as an abstract BST value and safely
used by pure code.  We are not aware of other functional programming
languages that can encapsulate destructive update in this way.

\begin{figure}
\begin{verbatim}
list_bst_du:: Ints -> BST
list_bst_du xs =
    *tp = Empty;                  -- allocate mem cell; init to Empty
    foldl_du bst_insert_du !tp xs -- repeatedly insert element

bst_insert_du tp x =              -- returns (), *tp updated
    case *tp of
    Empty ->
        *!tp := Node Empty x Empty  -- insert new node, return ()
    (Node *lp n *rp) ->
        if x <= n then
            (bst_insert_du !lp x) !tp -- update lp (and tp!)
        else
            (bst_insert_du !rp x) !tp -- update rp (and tp!)

foldl_du f y xs =                 -- returns (), y updated
    case xs of
    Nil -> ()                     -- return ()
    (Cons x xs1) ->
        f !y x;                   -- y updated by f
        foldl_du f !y xs1         -- y updated further
\end{verbatim}
\caption{BST creation using destructive update}
\label{fig_list_bst_du}
\end{figure}


In the second example we use another form of tree, for representing
cords.  Cords are data types which support similar operations to lists,
but concatenation can be done in constant time.  A common use involves
building a cord while traversing a data structure then converting the
cord into a list in $O(N)$ time, after which the cord is no longer used.
Here we use a simple cord design: a binary tree containing lists at
the leaves and no data in internal nodes.  Creating a cord from a list
plus append and prepend operations can all be done simply by applying
data constructors.

To convert such a cord to a list, a purely functional program would
typically copy each cons cell in each list.  A C programmer is likely
to consider the following more efficient algorithm, which destructively
concatenates all the lists without allocating any cons cells or copying
their contents.  For each list in the tree other than the rightmost
one, the \texttt{NULL} pointer at the end of the list is replaced with
a pointer to the first cell of the next list; the first list is then
returned (note this destroys the cord).  This algorithm can be coded
in Pawns -- see Figure \ref{fig_cord}.  The \verb@cord_list@ function
creates a pointer to an empty list and calls \verb@cord_list_a@, which
traverses the cord, updating this list (and the cord), then the list
is returned. \verb@cord_list_a@ is recursive and is always called with
a pointer to a \texttt{Nil}, which is updated with the concatenated
lists from the cord, and it returns a pointer to the \texttt{Nil} in
the updated list. For now we assume there are only lists of Ints (we
will briefly discuss impurity and polymorphism in Section
\ref{sec-type_safety}).


\begin{figure}
\begin{verbatim}
data Cord = Leaf Ints | Branch Cord Cord

-- convert list to cord
list_cord xs = Leaf xs

-- append two cords
cord_app xc1 xc2 = Branch xc1 xc2

-- append list to cord
cord_app_list xc xs = Branch xc (Leaf xs)

-- prepend list to cord
cord_prep_list xs xc = Branch (Leaf xs) xc

-- convert cord to list by efficiently smashing all the lists together -
-- what could possibly go wrong?...
cord_list xc =
    *xsp = Nil;                  -- pointer to empty list of Ints
    np = cord_list_a !xc !xsp;   -- smash all the lists together
    *xsp                         -- return (smashed) list

-- np points to Nil. We smash this list by appending all the lists in xc.
-- We return a ptr to the Nil at the end of the resulting list.
cord_list_a xc np =
    case xc of
    (Leaf xs) ->
        *!np := xs   !xc!xs;   -- smash Nil with xs
        lastp np               -- return ptr to Nil of updated np
    (Branch xc1 xc2) ->
        np1 = (cord_list_a !xc1 !np) !xc!xc2; -- append left subtree
        (cord_list_a !xc2 !np1)      !xc!np   -- append right subtree

-- returns pointer to the Nil of *xsp
lastp xsp =
    case *xsp of
    Nil -> xsp
    (Cons _ *xsp1) -> lastp xsp1
\end{verbatim}
\caption{Cord operations using destructive update}
\label{fig_cord}
\end{figure}

Compared to pure coding, this kind of coding is complicated and prone
to subtle bugs and assumptions (thus best avoided except where the added
efficiency is important).  It may seem that there are several redundent
``!" annotations but the Pawns compiler will complain without them.
For example, in the first recursive call to \verb@cord_list_a@,
with \texttt{xc1}, the compiler insists that \texttt{xc2} is
annotated. Although the analysis done by the compiler is unavoidably
conservative and sometimes results in false alarms, in this case it is
correct. It is possible the lists in the two branches of the cord may
share representations and if this is the case a cyclic list is created
and the code does not work! The same can occur if \verb@cord_list_a@ is
called with \texttt{xc} and \texttt{np} sharing, instead of \texttt{np}
pointing to an independent \texttt{Nil}.  The compiler insisting on extra
annotations hopefully alerts the programmer to these subtleties, leading
to better documentation and defensive coding to avoid the potential bug.

\section{Purity and abstraction}
\label{sec-pure-abs}

The distinction between pure and impure code can be blurred. For
example, some ``impure'' code can be given ``pure'' semantics by
introducing/renaming variables, adding function arguments \emph{et
cetera}. However, Pawns makes a different important distinction,
between data structures that are ``abstract'' (values for which the
representation is not important and may not be known) versus ``concrete''
(where the representation, including sharing, may be important and
should be understood by the programmer). Only concrete data structures
can be updated. Abstract data structures are normally associated with
pure code and concrete data structures with impure code but this is not
always the case.

Consider the \texttt{lastp} function of Figure \ref{fig_cord}. It takes
a pointer to a list, has no effects and always returns a pointer to
\texttt{Nil}, so in that sense it is pure (note that pointers themselves
are not impure).  However, for the destructive update code that uses
\texttt{lastp}, it matters \emph{which} \texttt{Nil} is pointed to in
the result. If \texttt{lastp} allocated a new memory cell, initialised it
to \texttt{Nil} and returned a pointer to this \texttt{Nil}, the result
would be identical from an abstract perspective but the \verb@cord_list@
code would not work.  Thus although \texttt{lastp} can be considered
pure, it must work with concrete data structures. Similarly, impure
functions can have abstract arguments and/or results (they cannot update
abstract arguments but may update other arguments).

When a data structure is created by applying a data constructor to
concrete arguments, the result is concrete. Concrete data structures can
become abstract when they are returned from a function (depending
on the type signature of the function) or if they are blended with
abstract data structures. For example, if the \texttt{Nil} of a
concrete list is updated with an abstract list or \texttt{Branch} is
applied to one or more abstract cords the result is abstract.  Pawns uses
the sharing system to keep track of the distinction between abstract and
concrete (see Section \ref{sec-sharing}).  Pure code such as that in
Figure \ref{fig_list_bst_pure} can be written without considering data
representation or sharing, but values returned from these functions will
be abstract and thus cannot be be updated.  Although \texttt{lastp} of
Figure \ref{fig_cord} is pure, the type signature must contain explicit
sharing information because we need a concrete list pointer to be returned
--- the representation is important and the data structure is intended to
be updated elsewhere.

\section{Sharing analysis}
\label{sec-sharing}

The Pawns compiler does \emph{sharing analysis} \cite{pawns-sharing}
to approximate how variables share components of their representations
and determine what variables may be updated at each point during evaluation
of each function \texttt{f}.  It relies on knowing what sharing may exist
between arguments in calls to \texttt{f}, what sharing may exist between
arguments and results of functions called by \texttt{f} and what arguments
of these functions may be updated. Type signatures in Pawns code have
additional information to help this analysis.  Specifically, they declare
which arguments may be updated, plus a ``precondition" stating what sharing
between arguments may be present when the function is called and a
``postcondition" stating what additional sharing may be present beween
arguments plus the result when the function returns. As well as the
compiler checking there are sufficient ``!" annotations, it checks that
whenever a function is called the precondition must be satisfied and when
a function returns the postcondition must be satisfied. Declaring this
additional information is a burden but it forces the programmer to think
about sharing in data structures that may be updated, documents sharing
for others reading or maintaining the code and helps the compiler conduct
analysis to check when destructive update can safely be encapsulated
inside pure code and used in the presence of polymorphism.
Preconditions can also be used to make code more robust.  For example, they
can be used to declare that no sharing should exist between the arguments
of \verb@cord_list_a@ or the functions that build cords: \verb@cord_app@,
\verb@cord_app_list@ and \verb@cord_prep_list@. Code where such sharing
exists will then result in a compiler error message instead of incorrect
runtime behaviour.

Abstract data structures share with a special pseudo-variable named
abstract (there are different versions of this variable for different
types \emph{et cetera}). For a function that contains no explicit
information concerning sharing, the default precondition and precondition
specify maximal possible sharing, including sharing with abstract. There
is no restriction on calls to such functions (preconditions are always
satisfied) but results share with abstract.  Code that attempts to
update a variable that shares with abstract results in a compiler error.
Similarly, passing an abstract data structure to a function that expects a
concrete data structure result in an error.  The data structure will share
with abstract but that will be at odds with precondition of the function.

A function that has no sharing declared \emph{can} return a concrete data
structure.  The implicit postcondition just specifies \emph{possible} (not
definite) sharing with abstract.  This allows code such as the definition
of \verb@list_bst_du@, where the interface is pure and abstract but the
implementation uses destructive update of a concrete data structure. As
a general rule in programming, if a possibly shared data structure is
updated, the programmer should understand how it has been used, all the
way back to the points where it was created.  In Pawns, this must be
documented in the code, by explicit declarations whenever it is passed
to or returned from a function, and these declarations are checked by the
compiler. At some later point we are free to treat it as an abstract value
and not concern ourselves with how it is represented or what may share
with it, but if this is done the value should not be updated further.
In Pawns, this is achieved by explicitly or implicitly adding sharing
with abstract.



Sharing is declared by augmenting type signatures with a pattern that
matches variables with the arguments and result of a function and pre-
and post-conditions that can use these variables. The pattern can also
prefix arguments by ``\texttt{!}'' to indicate the argument may be
updated.  Pre-conditions can use the arguments of the function (and
\texttt{abstract}) to declare the maximal sharing allowed when the
function is called.  Post-conditions can also use the result and
declare what additional sharing may be added during evaluation of
the function. The keyword \texttt{nosharing} is used to indicate no
sharing. Equations and other Pawns code (but not function calls) can be
used to indicate sharing between variables or components of variables
--- see Figure \ref{fig_sharing}.

\begin{figure}
\begin{verbatim}
list_bst_du:: Ints -> BST   -- explicit version of previous code
    sharing list_bst_du xs = t
    pre xs = abstract
    post t = abstract
bst_insert_du:: Ref BST -> Int -> ()
    sharing bst_insert_du !tp x = v
    pre nosharing
    post nosharing
foldl_du::
    ( Ref BST -> Int -> ()
        sharing f !xtp x = v
        pre nosharing
        post nosharing
    ) -> Ref BST -> Ints -> ()
    sharing foldl_du f !xtp1 xs = v
    pre nosharing
    post nosharing

lastp:: Ref Ints -> Ref Ints
    sharing lastp xsp = np
    pre nosharing
    post np = xsp
list_cord :: List -> Cord
    sharing list_cord xs = xc
    pre nosharing
    post xc = Leaf xs
cord_list:: Cord -> Ints
    sharing cord_list !xc = xs
    pre nosharing
    post xc = Leaf xs
cord_list_a:: Cord -> Ref Ints -> Ref Ints
    sharing cord_list_a !xc !np0 = np
    pre xc = Leaf *np0
    post np = np0
cord_app_list :: Cord -> List -> Cord
    sharing cord_app_list xc xs = xc1
    pre nosharing  -- If xs shares with lists in xc, list_cord breaks!
    post xc1 = inferred

\end{verbatim}
\caption{Type signatures with sharing}
\label{fig_sharing}
\end{figure} 

The declaration for \verb@list_bst_du@ here is equivalent to the
declaration in Figure \ref{fig_list_bst_du} but the sharing with
\texttt{abstract} is made explicit.  For the other \texttt{BST}
construction code there is no sharing. Integers are atomic; with a more
complex data type for elements there would generally be sharing between
the list and tree elements and this would need to be declared. Note that
even with no sharing, it needs to be declared, along with the fact that
the \texttt{BST} is updated, otherwise sharing with \texttt{abstract}
would be assumed and no update allowed. This applies equally to higher
order arguments such as that in \verb@foldl_du@.

The declarations for the cord code illustrate sharing of variables and
their components. Components of variables are discussed further below.
For \texttt{lastp}, the postcondition states that the result, \texttt{np},
and the argument, \texttt{xsp}, may be equal (and hence share all
components).  For \verb@list_cord@, the postcondition states the result,
\texttt{xc}, may be a \texttt{Leaf} whose argument is \texttt{xs}, the
argument of the function. This is exactly what the function returns but,
due to the imprecision discussed below, it means the argument of any
\texttt{Leaf} data constructor in \texttt{xc} may equal \texttt{xs}. This
more general interpretation is required for \verb@cord_list@.  Similarly,
for \verb@cord_list_a@, the precondition means a \texttt{Leaf} data
constructor argument of the cord may equal the list pointed to by
the second argument. The precondition of \verb@cord_app_list@ prevents
it introducing sharing between different lists in a cord, allowing the
compiler to reject code that has the bug mentioned earlier (the same
should be done for other cord construction functions). The postcondition
is inferred from the function definition --- this is supported in Pawns
for definitions that are pure and contain no function calls (potentially,
all postconditions could be inferred but we feel this would detract from
the philosophy of Pawns, which makes sharing obvious in the source code
wherever it must be understood by programmers).

Sharing analysis is unavoidably imprecise but it is conservative,
generally over-estimating the amount of sharing. Potentially, code may
need to have more sharing declared than is actually the case and more
variables annotated with ``!''. For each type, the sharing analysis uses a
domain that represents the memory cells that can be used for variables of
that type in the running program. For recursive types, the actual number
of memory cells can be unbounded, but ``type folding'' is used to reduce
it to a finite number. The domain distinguishes the different arguments
of different data constructors but where there is recursion in the type,
the potential nested components are all collapsed into one. For example,
for lists, there is a component for the head of the list and another for
the tail of the list but because lists are defined recursively, the head
component represents \emph{all} elements of the list (all memory cells
that are the first argument of a \texttt{Cons} in the list representation)
and the tail represents \emph{all} tails.

For cords, there are five components: the two arguments of
\texttt{Branch}, the argument of \texttt{Leaf} and the two arguments
of \texttt{Cons}.  Each left or right branch of a cord is a cord and
type folding makes the five components of the branches the same as the
top level cord.  Thus for \verb@cord_app_list@, the all five components
of \texttt{xc1} may share with the respective components of \texttt{xc},
along with the two components representing \texttt{Cons} arguments sharing
with the respective components of \texttt{xs}.  Sharing analysis keeps
track of what components may exist for each variable.  For example,
if a list variable is known to be \texttt{Nil} it has no components at
that point in the sharing analysis.  Also note that for two components
to share, they must have the same type and, unless they are pointers,
the same enclosing data constructor and argument. For example, the
argument of a \texttt{Leaf} cannot be the same memory location as the
second argument of a \texttt{Cons} and sharing analysis respects this
distinction. However, we can have a pointer that points to either of
these locations, thus sharing analysis treats pointers/refs differently
from other data constructors.

\section{IO and state variables}
\label{sec-state}

Like destructive update, IO does not fit easily with pure functional
programming.  Pawns models IO by using a value, representing the
state of the world, which is conceptually passed in and returned
from all computations that perform IO.  Rather than explicitly using
an extra argument and a tuple for results, \texttt{io} is declared as
``implicit'' in the type signature of functions (and nothing is actually
passed around). Pawns allows other ``state variables'' to be defined
and (conceptually) passed around in the same way.  In function type
signatures, they can be declared as ``ro'' (read only --- as if they
are passed in as an argument to the function), ``wo'' (write only ---
as if they are initialised/bound by the function and returned) or
``rw'' (read and written). The \texttt{io} state variable is bound
before the \texttt{main} function of a Pawns program is called and all
the primitive IO functions have \texttt{implicit rw io} in their type
signatures; other state variables must be explicitly bound/initialised
before being used. The state variable feature of Pawns is designed
so that pure functional semantics \emph{could} be defined. However,
calls to functions with implicit arguments/results must be prefixed
by \texttt{!} to highlight the fact than there is more going on in
the code than meets the eye, whether or not it is considered pure.
State variables are declared like type signatures of functions except
they are prefixed with \texttt{!} and must have a \texttt{Ref} type
(they point to a statically allocated memory cell and can be used for
destructive update like other pointers). They can only be used in code
after a \texttt{wo} function has been called or in functions where they
are declared implicit in the type signature.

\begin{figure}
\begin{verbatim}
bst_sum:: BST -> Int  -- sum of integers in a BST (pure interface)
bst_sum t =
    !init_nsum 0;   -- like nsum = 0
    !bst_sum_sv t;  -- like nsum' = bst_sum_sv t nsum
    *nsum           -- like nsum'

!nsum:: Ref Int  -- declares state variable, nsum

init_nsum:: Int -> ()
    implicit wo nsum  -- binds/initialises/writes nsum
init_nsum n =
    *nsum = n

bst_sum_sv:: BST -> () -- adds all integers in BST to nsum
    implicit rw nsum   -- reads and writes nsum
bst_sum_sv t =
    case t of
    Empty -> ()
    (Node l n r) ->
        *!nsum := *nsum + n;  -- adds n to nsum
        !bst_sum_sv l;        -- adds ints in l (could do same for r)
        *!nsum := *nsum + (bst_sum r) -- uses encapsulated impurity
\end{verbatim}
\caption{Summing the nodes in a BST using a state variable}
\label{fig_bst_sum}
\end{figure}

Figure \ref{fig_bst_sum} gives a simple example of summing the elements in
a BST using a state variable \texttt{nsum} instead of passing additional
arguments and results. Although \verb@bst_sum@ behaves as a pure function,
as the type signature implies, internally it uses \verb@init_nsum@ to
bind/initialise the state variable, which is updated as \verb@bst_sum_sv@
traverses the BST and then its final value is returned. State variables
are similar to mutable global variables in a language such as C but the
code makes it clear when the variables may be used/updated and they can
be encapsulated in a purely functional interface.  For example, although
\verb@bst_sum_sv@ calls \verb@bst_sum@ (which zeros \texttt{nsum} before
traversing the right subtree), Pawns ensures this does not interfere
with the \texttt{nsum} value in the outer computation.

Functions can have multiple state variables declared as implicit arguments
with no additional complications.  There is no ordering required for the
state variables, making some coding simpler compared to mechanisms other
languages use for threading multiple kinds of state in a pure way (such
as nested monads in Haskell). A disadvantage of using state variables is the
code is harder to re-use because it is tied to specific state variables
rather than types. State variables and their components can share and be
updated in the same way as other Pawns variables.  The only additional
restriction is that a state variable (or its alias) must not be passed to
code where the state variable is undefined (for example, be passed as an
argument or returned as a result of a function where the state variable
is not declared as an implicit argument). Thus \verb@bst_sum@ in Figure
\ref{fig_bst_sum} can return \verb@*nsum@ but not \texttt{nsum} itself,
even if the return type and/or the type of \texttt{nsum} was changed.

%
%
%
%


\section{Polymorphism and renaming}
\label{sec-rename}

Sharing in Pawns is not polymorphic to the same extent as
types. Similarly, code that uses a state variable is specific to that
state variable rather than something more general such as the monad type
class in Haskell.  For a function such as \texttt{foldl}, the second
and third arguments do not have identical types declared and Pawns does
not allow any sharing to be declared between them.  However, for some
calls to \texttt{foldl} the types may be identical and we may want to
declare sharing between them.  In Pawns, this can only be done by using
a separate function definition that has a more specific type signature
with identical types and the sharing declared. Pawns provides a mechanism
for renaming groups of functions to simplify this. As an example, Figure
\ref{fig_rename} shows how the code of Figure \ref{fig_list_bst_pure}
code can be duplicated, making it possible to add different type
signatures where the sharing is declared and hence the resulting
tree can be updated\footnote{There is little advantage in having both
abstract and concrete versions of these functions but it does illustrate
renaming}.  The first \texttt{renaming} declaration creates definitions
of \verb@list_bst_concrete@ and \verb@bst_insert_concrete@, by renamining
the previous definitions and replacing the call to \texttt{foldl} by a
call to \texttt{foldlBST}. An explicit definition of \texttt{foldlBST}
could be included but we here simply use another \texttt{renaming}
declaration. Type signatures are needed for all three functions (for
brevity we just include one).  Renaming can also be used as a less
abstract alternative to higher order code and for producing code with
the same structure but with different state variables.  For example,
we can code a version of \texttt{map} that uses \texttt{io} and rename
it to use other state variables as needed (this is the Pawns equivalent
of using Haskell's \texttt{mapM}).


\begin{figure}
\begin{verbatim}
renaming
    list_bst_concrete = list_bst_pure
    bst_insert_concrete = bst_insert_pure
    with
    foldlBST = foldl

-- same effect as just deleting the "with" above
renaming
    foldlBST = foldl

-- also need type signatures for list_bst_concrete and foldlBST
bst_insert_concrete:: BST -> Int -> BST
    sharing bst_insert_concrete xt x = xt1
    pre nosharing
    post xt1 = xt
\end{verbatim}

%
%
\caption{Renaming of function definitions}
\label{fig_rename}
\end{figure}

\section{Complications}
\label{sec-comp}

Combining pure functional programming with destructive update and other
impurity is not simple!  The design of Pawns aims to support high
level pure functional programming plus low level imperative programming
with as much flexibility as possible while avoiding unsafe operations
(such as dereferencing \texttt{NULL} pointers) and ``surprises'' (code
with effects that are obscure). Here we briefly mention some of more
complicated issues and how they are dealt with in Pawns, without too
much technical detail.

\subsection{Polymorphism and type safety}
\label{sec-type_safety}

Mixing polymorphic types with destructive update can result in unsafe
operations if it is not done carefully. Consider the code in Figure
\ref{fig_type_safety}. The variable \texttt{xsp} is bound to a pointer
to \texttt{Nil}, a list of \emph{any} type.  Without destructive
update, this can be safely used where pointers to lists of integers
and pointers to lists of binary search trees are expected (the type
can be instantiated to either of these without problems). However,
if the variable is updated to be a non-empty list of integers the code
is not type safe --- an integer may appear where a tree is expected.
Other functional languages solve the type safety problem by imposing
restrictions on code that has refs (and thus may perform updates). In
Pawns, refs to arguments of data constructors can be created anywhere,
but because the source code explicitly notes where variables can be
updated, the problem can be solved in a more flexible way.

Where a Pawns variable with a polymorphic type is assigned to or passed to
a function that may update it, type variables may become more instantiated
during type checking.  For example, at the point where \texttt{xsp} is
passed to \verb@int_fn@ in Figure \ref{fig_type_safety}, its previous
polymorphic type (\texttt{Ref (List t)}) is instantiated to \texttt{Ref
(List Int)}.  The type of \texttt{xsp1} is also similarly instantiated ---
the two variables share their representations and their types shared the
same type variable, \texttt{t}.  The subsequent call to \verb@bst_fn@
then results in a type error. Pawns treats all variables created with
polymorphic types as live throughout the whole function, so the \texttt{!}
annotation on \texttt{xsp} is required even if \texttt{xsp} is never used
again, alerting readers of the source code to a subtlety. The compiler
also prints a warning when types are further instantiated. Warnings can
be avoided by adding explicit casts, as shown in the second example of
Figure \ref{fig_type_safety} (a previous version of the compiler did
not automatically instantiate types and this cast was required).

\begin{figure}
\begin{verbatim}
    *xsp = Nil;       -- Nil is a list of any type
    xsp1 = xsp;       -- xsp1 has the same polymorphic type as xsp
    ys = (int_fn !xsp) !xsp1; -- int_fn accepts a ref to list of ints
    -- Now *xsp (and *xsp1) may be a non-empty list of ints!
    zs = bst_fn *xsp1; -- OOPS! bst_fn accepts a ref to list of BSTs

cord_list xc =
    *xsp = Nil::Ints;           -- instantiate list type explicitly
    np = cord_list_a !xc !xsp;  -- xsp has a monomorphic type
    *xsp
\end{verbatim}
\caption{Potential violation of type safety}
\label{fig_type_safety}
\end{figure}

\subsection{Higher order programming}

There are two complications involving higher order code: type checking
and partially applied functions (closures).  Type checking is made
more complicated because each ``arrow'' type has additional information
concerning sharing, destructive update and state variables. Pawns allows
some latitude when matching the type of arguments to higher order
functions with the expected type that is declared. The arguments are
allowed to have less destructive update, less sharing in postconditions,
more sharing in preconditions and some variations in what state
variable operations are declared (for example, \texttt{ro} is acceptable
where \texttt{rw} is declared). The intention is to allow as much
flexibility as possible while guaranteeing safety.

Pawns allows functions to be applied to fewer than the declared number of
arguments, resulting in closures being constructed/returned. Closures
can be passed around like other data and later applied, leading to
function evaluation. The arguments inside closures can share with other
data structures and hence they can potentially be updated. Pawns allows
the patterns used for declaring sharing to have additional arguments,
representing the arguments of closures, so sharing of data within closures
can be declared and analysed.  Certain equivalence laws that hold for
pure functional programming (such as ``eta-equivalence'') do not apply
when sharing is significant and there may be destructive update.

\begin{figure}
\begin{verbatim}
put_char: Int -> ()
    implicit rw io
put_char i = as_C "{putchar((int) i);}"

-- pseudo-random number sequence interface
init_random:: int -> () -- initialize sequence with a seed
    implicit wo random_state
random_num:: () -> int  -- return next number in sequence
    implicit rw random_state
\end{verbatim}
\caption{C interface}
\label{fig_C}
\end{figure}

\subsection{Foreign language interface}

The one feature of Pawns where there is no attempt to guarantee safety
is the foreign language interface. Pawns compiles to C and provides
a simple and flexible interface to C, which has many unsafe features.
Each Pawns function compiles to a C function and Pawns allows the body
of a function definition to be coded in C but for such code there
can be no guarantees of safety or lack of ``surprises''.  It is up
to the programmer to ensure the C code is safe and compatible with
the Pawns type signature.  For example, Figure \ref{fig_C} gives the
implementation of \verb@put_char@ defined in terms of \texttt{putchar}
in C. The use of the \texttt{io} state variable in the type signature
ensures that the code can only be used in a context where the side-effect
is clear and purely functional semantics could be defined.  Similarly,
it only requires a few lines of code to interface Pawns to the C standard
library pseudo-random number package in a way that can be encapsulated
and given purely functional semantics, using a state variable --- see
Figure \ref{fig_C} for the type signatures.  It is also very easy to
support arrays via the C interface; the current code has no bound checks
(and thus has C-like efficiency but is not safe).

Most foreign language interfaces only allow basic unstructured types to
be passed.  However, the Pawns compiler uses the \texttt{adtpp} tool
\cite{adtpp}, which generates C macros for manipulating the algebraic
data types defined in the program.  For example, Pawns code that defines
the \texttt{BST} type results in C macros for creating an \texttt{Empty}
tree, creating a \texttt{Node} and various ways of testing if a tree
is \texttt{Empty} or a \texttt{Node} and extracting the arguments of
the \texttt{Node}.  These macros can be used in hand-written C code
to both operate on a \texttt{BST} that was created by Pawns code, and
create a \texttt{BST} that is passed back to Pawns code.  Dynamic memory
management is often particularly difficult across language boundaries but
is made very easy in Pawns by using the Boehm-Demers-Weiser conservative
garbage collector.

\section{Conclusion}
\label{sec-conc}

There are important algorithms which rely on destructive update of
shared data structures, and these algorithms are relatively difficult to
express in declarative languages and are typically relatively inefficient.
The design of Pawns attempts to overcome this limitation while retaining
many of the advantages of a typical functional programming language,
such as algebraic data types, parametric polymorphism, and higher order
programming.
Pawns supports the creation of pointers to arguments of data constructors,
and these pointers can be used for destructive update of
shared data structures.  There are several features which restrict
when these effects can occur and allow them
to be encapsulated, so the abstract declarative view of some functions
can still be used, even when they use destructive update internally.

Type signatures of functions declare which arguments are mutable and
for function calls and other statements, variables are annotated if it
is possible that they could be updated at that point.  In order to
determine which variables could be updated, it is necessary to know what
sharing there is.  Functions have pre- and post-conditions which
describe the sharing of arguments and the result when the function is
called and when it returns.  To avoid having to consider sharing of data
structures for all the code, some function arguments and results can be
declared abstract (this is the default).  Reasoning about code which only
uses abstract data structures can be identical to reasoning about pure
functional code, as destructive update is prevented.
Where data structures are not abstract, lower level reasoning must be
used --- the programmer must consider how values are represented and
what sharing exists.  The compiler checks that declarations and
definitions are consistent, allowing low level code to be safely
encapsulated inside a pure interface.  Likewise, the state variable
mechanism allows a pure view of what are essentially mutable global
variables, avoiding the need for source code to explicitly give
arguments to and extract result from function calls. Analysis of
sharing is also required to ensure the use of state variables can be
encapsulated and to ensure safety of code that uses destructive update
of polymorphic data types.

Although Pawns is still essentially a prototype, and is
unlikely to reach full maturity as a ``serious'' programming language,
we feel its novel features add to the programming language landscape.
They may influence other languages and help combine the declarative and
imperative paradigms, allowing both high level reasoning for most code and
the efficiency benefits of destructive update of shared data structures.

\section*{Acknowledgements}
The design of Pawns has benefitted from discussions with many people.
Bernie Pope and Peter Schachte particularly deserve a mention.

\bibliography{all}
%
%
%
\end{document}